# Trusting AI in High-stake Decision Making – Ali Saffarini

Since the release of ChatGPT, people quickly became accepting of having artificial intelligence models draft emails asking for deadline extensions for them, summarizing information about certain subject matters for them, or even writing code for them. But what if it came to having AI choose what university we should attend, or which job offer we should sign? Arvai et al. (2021) from the University of Michigan's Erb Institute concluded in a study that when it came to financial and healthcare decisions, "participants indicated a strong preference for advice from human experts over AI." It appears that when it comes to making these high-stakes decisions, artificial intelligence models are no longer our first advisors. Despite artificial intelligence's quantitative, evaluative, and analytical capabilities, there are certain elements it lacks that lead humans to not feel so confident about the systems' advice or decisions. This creates an interesting gap of exploration – what would need to happen for humans to trust artificial intelligence models for important decisions? Such an exploration is significant because the conclusions drawn from it have broader implications on humans' control of society in terms of healthcare, governance, and more areas where decisions have ripple effects on society.

In the discussion of trusting artificial intelligence for important decisions, researchers and developers seem to converge their focus on enhancing explainability and interpretability, thus increasing people's trust in the system. Experts like Daniela Rus (2019) from MIT, Birger Thorburn (2019), and others explain how explainability is a crucial component of ensuring trust within these intelligence systems, and they elicit the role of explainable artificial intelligence in augmenting the level of people's trust within systems. Yet, this technical approach to what needs to happen for humans to trust artificial intelligence in decision-making misses out on the more tangible aspect of trust – the psychological aspect. Certain psychological trust barriers come from our interaction with the intelligence models and to create AI models that can be trusted, there should be a heavier consideration of psychological aspects of mistrust in AI. By this, the models can be technically modified so that they overcome these psychological barriers rather than overcoming just the technical flaws of the system. Trust in artificial intelligence to make important decisions will be accomplished if developers anthropomorphize the models, modifying them in ways that address the psychological barriers of trust.

    In the context of this discussion, trust refers to our perception of the system's ability to give us correct information, justifiable evaluation and synthesis, and proper execution of tasks. The development of this trust between a human and a system is accomplished in multiple stages. Mischa Coster (2023) elaborates on this. One of the stages is dependability-based. This trust is established based on the system's performance with the tasks it is given. Interestingly, when an AI system makes a mistake, "trust in it decreases more than if a human had made the same mistake." (Coster, 2023) The reverse is true – when the model is successful, "trust in it increases more than if a human succeeded." (Coster, 2023) For this, consistency and reliability are huge factors to enforce trust within a system. Another stage also explained by Coster (2023) is "predictability-based trust." The more the user interacts with the system, their trust becomes proportional to the predictability of the system. If our anticipation of the system's output is aligned with the system's actual output, we lay more trust in the system. (Coster, 2023) Research from the University of Toronto's Graduate School of Computer Science explains how "techniques from explainability and interpretability are essential for building models whose behaviour can be anticipated by a user." (Madras, 2022, p. 58) For this reason, explainability and transparency are influential factors when it comes to trust.

Because consistency, reliability, and explainability are such key factors in creating trusted artificial intelligence models, experts in the field focus on solving these issues technically for humans to trust artificial intelligence even for the most high-stakes decisions. Department chair for Cyber Intelligence and Data Science, Mark Bailey (2023) talks about the main difficulty associated with all these traits of trust. The reality is artificial intelligence models are fundamentally unexplainable in that they cannot rationalize their decisions. This inability of artificial intelligence models to explain their rationale is termed the black box problem. Essentially, this is the idea that artificial intelligence systems work in the ways they're supposed to, we just don't understand how these models draw their conclusions. We are unable to see how models mathematically function to make their decisions. (Rawashdeh, 2023, as cited in Blouin, 2023) Because the black-box problem has such negative implications on trusting artificial intelligence, experts in the field are trying to solve it. Explainable artificial intelligence (XAI) is one of the evolving technologies that aims to help users understand how decisions are made. (Dee, 2023) In high-stakes domains, the presence of explainable models would give users a better understanding of the workings of these artificial intelligence systems, hence increasing trust within the system. (Markus, 2021) Most of the focus on making artificial intelligence more trusted is placed on altering the technology to partially resolve the black-box problem through XAI.

These technical considerations to enhance trust within artificial intelligence models may not enhance trust as much as anticipated. To start, explainable artificial intelligence as a solution isn't all that promising. Up until now, explainable artificial intelligence models do not accomplish all the objectives they should be according to Jessica Newman from UC Berkeley (2021). In addition, according to Thorburn (2019), "fully explainable AI systems are still a utopia". This incompleteness of explainable artificial intelligence creates an inability to make people trust artificial intelligence systems. From another lens, even in an idealistic world where explainable systems exist and functioned the way they should be, Heather Roff (2023, as cited in Anderson and Rainie, 2023), a nonresident fellow at the Brooking institutions and senior research scientist at the University of Colorado-Boulder states that "Most users are just not that fluent in AI or how autonomous systems utilizing AI work, and they don't care." For example, not all drivers know how their car functions, but they still trust it as a machine and drive it daily. When it comes to high-stakes decision-making, explanations provided by the model are also subject to questioning. Should we trust the rationale that induced the model's decision? The answer to that question is that we don't really need to answer it because XAI is not technically likely. After all, artificial intelligence models don't understand words, they deal with numbers and mathematical functions that understand relations between words, not their true concepts. By this we conclude that tackling what needs to happen for users to trust artificial intelligence solely from the technological perspective of the black-box problem is flawed; even if we had robust explanations, we wouldn't fully entrust the systems. Moreover, this idea that the opacity of artificial intelligence systems is one of the main reasons for users not trusting is also challenged by the idea that the human mind is opaque as well. Not only do scientists not fully understand how the human mind works, but in addition, we don't always understand how and why some people think in the ways they do. (Freitas et al., 2023) From this it's concluded that other trust barriers need to be addressed, in tandem with these technical ones, for trust in AI for important decisions to be accomplished.

The other main aspects of the struggle to establish trust between artificial intelligence systems and humans are behavioral and psychological. John Blythe (2023) explains how "unlike humans, AI lacks consciousness, emotions, and personal accountability." The absence of human-like traits within systems creates difficulty in trusting them because we essentially treat them in ways that are different from the ways we treat humans. This brings about the psychological barriers to trusting artificial intelligence models. It turns out that the issue of mistrust isn't just because of the explainability of the technology, it's because of natural human behavior in the way they interact with systems and entities. By examining these behavioral trust barriers, we could synthesize how we can alter the models so that they overcome these barriers and thus lead users to trust them.

The psychological barriers to trust in artificial intelligence stem from the system being emotionless, inflexible, and autonomous. Freitas et al. (2023) explain how these psychological factors influence our trust in these systems. When it comes to important decision making, artificial intelligence models miss out on a key factor which is that they don't experience emotions. "Emotional abilities are associated with subjective tasks" (Freitas et al., 2023) and most important decisions have some level of subjectivity within them. Some emotional considerations need to be considered when making an important decision, and the artificial intelligence model should be able to reflect that in the way it transmits output. The issue here is that artificial intelligence models are perceived as very impartial, and for that, we don't react positively to the outputs (or their explanations if any) even if the system did quantify and evaluate the same emotional considerations that any human would. The other factor is inflexibility. The psychological analysis here is that we are subconsciously accustomed to software being rigorous – they'll always function in the same way. So, if a system mistakes something once, it will consistently do so. This is what we anticipate when AI makes a mistake, and thus our trust in it decreases significantly more. Our historical understanding of machines being inflexible (they operate in the same way repetitively) influences our recognition that machine learning can learn from its mistakes. (Freitas et al., 2023) Also, because people naturally "prefer to take actions that give them more choice rather than less," (Freitas et al., 2023) people will likely not adopt artificial intelligence models that essentially inflict people's freedom and choice in their actions. This factor of autonomy relates to humans' reasons for resisting change. (Kanter, 2012) Humans, for extended periods of time, have been relying on human experts to make important decisions for them. Because artificial intelligence is still a new concept, people are still reluctant to make this huge switch; our reliance on humans for important tasks has been working well. Once again, it's important to elicit here that this analysis is truer for important tasks or decisions than less important ones. For example, people won't resist using artificial intelligence robots to clean or sweep their homes, but they are more likely to resist using artificial intelligence robots to diagnose them or other tasks that have more severe consequences. One other psychological barrier to trusting AI is that AI simply hasn't been here long enough for us to trust it. Research by Bach et al. (2022) shows that the more frequently we use AI, the more we tend to trust it. These psychological barriers are strongly preventive of humans trusting AI to make important decisions.

     Studies conducted by a Google researcher, Dr. Vyacheslav Polonski (2018), showcase the psychological barriers when it comes to trusting artificial intelligence systems, and from them conclusions about how we can overcome these barriers can be extracted. IBM developed an artificial intelligence decision system called Watson for Oncology; this system would advise doctors on the treatment of 12 cancers. (The Conversation US & Polonski, 2018) The study

shows that the doctors simply didn't trust the system if it gave different treatment recommendations than the doctors. "IBM has yet to provide evidence that Watson improves cancer survival rates." (The Conversation US & Polonski, 2018) This exemplifies how artificial intelligence hasn't been used for long enough to prove its credibility. Additionally, even if Watson was able to explain its decisions (the black-box problem), experts would enter debates between the AI and their explanations and justifications. A Danish hospital using the system had its doctors disagreeing with it in over two-thirds of the cases. (The Conversation US & Polonski, 2018) Because doctors were entering these medical disputes and because humans naturally prefer autonomy as explained before, the doctors would simply put their judgment at the front. The healthcare institutions that were using Watson ended up dropping them. The study reinforces the idea that humans are psychologically resistant to trusting these intelligence models. The study helps us conclude that one way to increase trust within the systems is by involving people more in the process. By giving people the freedom to slightly modify algorithms or intelligence models, we'd likely start to believe and trust that source. (The Conversation US & Polonski, 2018) This realization that there must be a connection between users and the system to overcome the psychological barrier of trusting artificial intelligence is significant because it steers our focus toward a more sensitive topic to enhance trust.

By modifying artificial intelligence models to evoke certain human characteristics, humans are inclined to trust it. A recent study at Stanford University's Human-Centered Artificial Intelligence showed that people are more likely to trust artificial intelligence if it shows or evokes certain characteristics or emotions. (Hadhazy, 2022) The study showed that when certain artificial intelligence models generated output with exaggerated confidence levels, users were more confident in the output (in this case advice) given by the artificial intelligence model.

In other words, by evoking the human trait of confidence and assertation the participants were more likely to trust the artificial intelligence model. From this, it is concluded that for a system to be impartial, which some may term as minimal character, it would likely be difficult to trust. Impartiality evokes doubtfulness whereas modifying the system to sound more confident stimulates trust. In another study related to self-driving cars conducted by a psychologist at Northwestern University, Adam Waytz, it was found that people were less startled by a staged car crash in a self-driving car when the self-driving car had a human name (Iris) and spoke in a female voice in comparison to a no-name vehicle. (Hutson, 2017) These results suggest "that they [participants] trusted the car because they had anthropomorphized it." (Hutson, 2017) The results of both these studies bring us to an astounding deduction.

    For humans to overcome the psychological barriers and trust artificial intelligence systems, the models would need to be anthropomorphized - seem more human. The studies explained above show that humans tend to connect more to other humans, or, in this case, human-like systems and we've previously concluded that a connection between the user and a system is essential. Trust is a form of connection, and if users are somewhat deceived that artificial intelligence models are human-like, we are more likely to trust them. In more analogous language, anthropomorphizing artificial intelligence models is like giving them a disguise – this gives two important keys to humans gaining trust. The first key is that the abrupt transition into believing a non-human is more discrete because the models behave in ways similar to ours. The second key is that by disguising these models, we'll gradually realize the actual competence of the systems. As explained previously, usage of the systems and time is required for it to prove its credibility. Moreover, the anthropomorphism of artificial intelligence overcomes the psychological barriers previously explained. First, anthropomorphism overcomes the barrier of

being emotionless because the very first step of making systems human is by changing their output style (be it tone, language, etc.) to evoke certain characteristics and emotions. Making models sound more confident, sympathetic, kind, or even stubborn would reinforce our subconscious perception of the system as one that can recognize emotion, and thus consider emotion in high-stake decision-making. Other than that, by perceiving machines as more human, we immediately disassociate the thought of inflexibility from the system. Humans are very quick at adapting, changing, and learning. Thus, any mistakes by the programmed machines would seem less of an issue to us because we anticipate that the machine would be able to learn – and they can. Prior to anthropomorphizing the machine, we stuck to our notion that if a machine is wrong once, it'll continue to be unless fixed because that is our subconscious historical understanding of software and machines. Lastly, anthropomorphism would overcome the barrier of humans preferring autonomy because for years humans have delegated authority to other humans – such as voting or even hiring managers for companies. If our perception of machines is that they are more *human*, then we will likely choose to delegate authority to them because our minds would observe that they behave like us as they are anthropomorphized. By that, the anthropomorphism of artificial intelligence systems would enable humans to trust these systems, even for important decisions.

  There are important considerations that need to be considered alongside this claim of anthropomorphized AI would accomplish our trust in the systems for high-stake decisions. First, anthropomorphism is not going to necessarily make AI more trustworthy. There is a distinction between trust and trustworthiness; people can trust non-trustworthy entities and vice versa. The idea here is that developing explainable artificial intelligence and getting closer to solving the black-box problem deals with enhancing the trustworthiness of AI models. Even with a solution to the black-box and therefore a trustworthy AI model, people may still not trust it due to psychological factors. Whether trustworthy or untrustworthy, artificial intelligence models must appeal to humans in some way for them to trust it. Anthropomorphism offers that appeal; it is deceitful in that it gives the impression that it is similar to what humans have been relying on – themselves. Because of this distinction, it is extremely important to recognize the importance of not putting a hold on the tackling of the technical difficulties of AI models to enhance trustworthiness. If developers wish to anthropomorphize AI to make people trust it, they must improve the explainability of AI to improve its trustworthiness as well. One other consideration is that it can be argued that there is no need to anthropomorphize artificial intelligence systems to trust them; we just need to use them long enough for us to realize their competence in making important decisions. There is a flaw in this counterargument because to reach such a point of using AI for long enough, users must be given reason to trust it initially; we need to trust the AI to begin using it so it can be used long enough for us to further trust its capabilities. This initial trust comes from anthropomorphism.

      To conclude, the anthropomorphism of artificial intelligence technology is a necessary modification for humans to trust artificial intelligence even for high-stake decisions because this modification addresses the psychological barriers of trust. Our research is a recognition that even with exponential advancements in technology that could make artificial intelligence systems reliable and consistent (trustworthy), our reluctance to give away autonomy, change, and positively interact with emotionless systems hinders the establishment of a connection of trust. Anthropomorphism here is making our perception of the system more human. This can occur in the form of the writing style, voice, or even appearance if it were physical robots. This process would enable a more natural connection between AI and humans,

reaching a point where we'd be confident in the system's abilities to make important decisions for us. It's important to resume the development of AI technology so that if we do end up trusting these systems, we trust ones of good fidelity and reliability. Yet this very idea of making non-human systems seem more human is eerie; the process of anthropomorphism may entail many implications for society and technology. While one of these implications as concluded from this paper is enabling humans to trust artificial intelligence models for important decisions, other implications could bring either or both benefits and dire consequences to society. This makes for both an interesting and crucial area of exploration because if developers do decide to anthropomorphize AI to enhance users' trust in it for important tasks and decisions, we should be aware of its broader implications.